\documentclass[aps,preprint,nofootinbib,floatfix,superscriptaddress]{revtex4}
\usepackage{epsfig}
\usepackage{graphicx}
\usepackage{multirow}
\usepackage{latexsym,amsbsy,amsmath}
\usepackage{stackrel}
\def\ll{\left\langle}
\def\rr{\right\rangle}

\def\rp{\right|}
\def\r.{\right.}
\def\l.{\left.}
\begin{document} 
\title{\bf $\Omega_{bbb}\Omega_{bbb}\Omega_{bbb}$ tribaryons} 
\author{H.~ Garcilazo}
\email{hgarcilazos@ipn.mx} 
\affiliation{Escuela Superior de F\' \i sica y Matem\'aticas \\ 
Instituto Polit\'ecnico Nacional, Edificio 9,
07738 CDMX, Mexico}
\author{A.~ Valcarce}
\email{valcarce@usal.es} 
\affiliation{Departamento de F\'isica Fundamental,\\ 
Universidad de Salamanca, E-37008 Salamanca, Spain}
\date{\today} 

\begin{abstract} 
We study the possible existence of bound states of three $\Omega_{bbb}$
baryons. We consider only $S$ wave interactions and we start from
recent lattice QCD results which give a strongly attractive potential between
two $\Omega_{bbb}$ baryons in the $^1S_0$ channel. We analyze 
different scenarios. At baryonic level, the
$\Omega_{bbb}\Omega_{bbb}$ interaction could be understood to be
basically spin-independent, so that the two contributing channels, 
$^1S_0$ and $^5S_2$, would have a very similar interaction.
This baryonic analysis leads to the existence of bound states 
in the three-body system. At the quark level, repulsive effects
would appear in the $^5S_2$ channel, making it more repulsive than 
the $^1S_0$ channel. We study the effect of such repulsion in terms 
of its range.
\end{abstract}
\maketitle 
 
\section{Introduction}
\label{sec1}
Recently, there have been several interesting
developements on the possible existence of  bound states of two and three 
$\Omega$ baryons. For example, Ref.~\cite{MING}, using a one-boson-exchange (OBE)
model found bound states of the systems 
$\Omega_{ccc}\Omega_{ccc}$ and $\Omega_{bbb}\Omega_{bbb}$. 
Ref.~\cite{GONG} derived a $\Omega\Omega$ interaction
based on lattice QCD. Similarly, Ref.~\cite{YANL},
a lattice QCD calculation with nearly physical light-quark masses,
derived the $\Omega_{ccc}\Omega_{ccc}$ interaction in the $^1S_0$ 
channel. They obtain a bound state with a binding energy of 5.68 MeV.
More recently, Ref.~\cite{MATH} performed a lattice QCD calculation of the 
$\Omega_{bbb}\Omega_{bbb}$ system finding a very deep  bound state 
in the $^1S_0$ channel, with a binding energy of 81 MeV.
The energy of the bound states of the two-body systems would be the
threshold of any possible three-body bound state. Finally, 
Ref.~\cite{TIAN}, using the existing lattice QCD interactions
for the different $\Omega\Omega$ systems, investigated the three-body systems
$\Omega\Omega\Omega$, $\Omega_{ccc}\Omega_{ccc}\Omega_{ccc}$, and
$\Omega_{bbb}\Omega_{bbb}\Omega_{bbb}$. They found that none of the three-body
systems binds. However, making use of the OBE interactions
of Ref.~\cite{MING} the $\Omega\Omega\Omega$ system develops a
bound state.

In this work, we investigate whether the 
$\Omega_{bbb}\Omega_{bbb}\Omega_{bbb}$ system is bound.
The $\Omega_{bbb}$ baryon has spin $3/2$ and no isospin, so that the two-body
system can have total spin $S_i=0, 1, 2$, and $3$. However, the states $S_i=1$
and $S_i=3$ are not allowed in $S$-wave by the Pauli principle, so that one 
is left with only the states $S_i=0$ and $S_i=2$. 
In Ref.~\cite{MATH}, they obtained the $\Omega_{bbb}\Omega_{bbb}$
interaction only for the channel $S_i=0$, so that we will have to discuss the
situation of the channel $S_i=2$. We will use
some hypotheses deduces either at the 
baryon level or at the quark level about the 
$\Omega_{bbb}\Omega_{bbb}$ $^5S_2$ interaction.

We carry out our study within the formalism of the 
nonrelativistic Faddeev equations for
three identical particles considering only $S$ waves.  
We start our discussion of the three-body system considering 
only the $S_i=0$ two-body channel and afterwards we analyze 
the effect of the $S_i=2$ two-body channel for the three-body bound state.

\section{Single channel Faddeev Problem}
\label{sec2}
The Faddeev equations for three identical particles are
\begin{equation} 
T=2t_iG_0T \,\, ,
\label{eq1} 
\end{equation}
where $t_i$ is the $t$-matrix of the two-body system,
\begin{equation} 
t_i=V+VG_0t_i \,\, ,
\label{eq2} 
\end{equation} 
where $V$ the two-body interaction in the $^1S_0$ channel and $G_0$ is
the propagator for three free particles.

We use the complete set of basis states $\left| i \right\rangle$,
\begin{equation} 
\left| i \right\rangle = \left| p_iq_i((s_j,s_k)S_i,s_i)J \right\rangle \,\, ,
\label{eq3}
\end{equation}
with $p_i$ and $q_i$ the standard Jacobi momenta, $s_i$, $s_j$, and $s_k$ the
spins of the three particles, $S_i$ the total spin of the pair $jk$, and $J$
the total spin of the three-body system. In this basis the Faddeev 
equation~\eqref{eq1} becomes,
\begin{equation} 
\left\langle i \right| T \left| \phi_0 \right\rangle =
2 \left\langle i \right| t_i \left| i'\right\rangle
\left\langle i'\right|\left. j \right\rangle G_0 \left\rangle j 
\right| T \left|\phi_0 \right\rangle \,\, ,
\label{eq4}
\end{equation}
where the explicit form of the integral equation in momentum space is given
in the Appendix.

The recoupling coefficient,
\begin{equation} 
\ll i'\rp \l. j \rr = \ll p_i^\prime q_i^\prime \rp \l. p_jq_j \rr 
\ll ((s_j,s_k)S_i,s_i)J \rp \l. ((s_k,s_i)S_j,s_j)J\rr \,\, ,
\end{equation}
is of great interest. The space part $\ll p_i^\prime q_i^\prime \rp \l. p_jq_j \rr$ 
is positive definite~\cite{BALI}; however, the spin part is
\begin{equation} 
\ll ((s_j,s_k)S_i,s_i)J \rp \l. ((s_k,s_i)S_j,s_j)J \rr = 
(-)^{S_j+s_j-J}\sqrt{(2S_i+1)(2S_j+1)} W(s_js_kJs_i;S_iS_j) \,\, .
\label{eq6}
\end{equation} 
Since $S_i=S_j=0$ and $s_i=s_j=s_k=J=3/2$ one gets,
\begin{equation} 
\ll ((s_j,s_k)S_i,s_i)J \rp \l. ((s_k,s_i)S_j,s_j)J \rr =
(-)^{2s_j}\frac{1}{2s_j+1} = -\frac{1}{4} \,\, ,
\label{eq7}
\end{equation}
which is a negative number, so that it effectively changes the nature of the
two-body interaction from attractive to repulsive such that no bound state can
be obtained in a one-channel calculation.~\footnote{On the opposite side, a repulsive
potential together with a negative spin recoupling coefficient does not change
the nature of the interaction so as to lead to a three-body bound state, as
discussed in Ref.~\cite{GARC} for the case of the $J=3/2$ $\Sigma^-nn$ system.}
    
The result of Eq.~\eqref{eq7} is a direct consequence of the Pauli principle and it
applies for all systems with three identical fermions, i.e., particles with
spin half-integer, like the case of three neutrons where,
\begin{equation}  
\ll ((s_j,s_k)S_i,s_i)J \rp \l. ((s_k,s_i)S_j,s_j)J \rr = 
(-)^{2s_j}\frac{1}{2s_j+1} =-\frac{1}{2} \,\, .
\end{equation}
It is worth noting that in the three-neutron case there is only a
two-body channel, $S_i=0$, if one includes only $S$ waves, so that there is no 
possibility for a three-neutron bound state with any interaction.
However, in the case of the three omegas besides the $S_i=0$ channel
one also has the $S_i=2$ two-body channel.

\section{Baryonic level $^5S_2$ $\Omega\Omega$ interaction}
\label{sec3}
A two-body interaction acting in $S$-waves contains only central and spin-spin
terms, since terms like spin-orbit, tensor, etc., act only for nonzero orbital
angular momentum.

The phenomenological description of the spectra of mesons and baryons
in a nonrelativistic approach is based in a two-body 
potential between quarks~\cite{BHAD,BRAC}. In Ref.~\cite{BHAD} 
such potential was taken to be, 
\begin{equation} 
V(r)=-\frac{\kappa}{r}+\lambda r-\Lambda+\frac{\kappa}{m_im_j}
\frac{exp^{-r/r_0}}{rr_0}\vec\sigma_i\cdot\vec\sigma_j
\label{eq8}
\end{equation} 
and similarly in Ref.~\cite{BRAC}. The four terms in the r.h.s. of
Eq.~\eqref{eq8} are, respectively, the Coulomb term, the linear confinement 
term, the constant term, and the spin-spin term. This interaction is able to
reproduce reasonably well the masses and other properties of all the existing
mesons and baryons~\cite{BRAC}.

The Yukawa function in the spin-spin term is an extended delta function which
becomes a delta function if $r_0\to 0$. This form of the spin-spin interaction
is suggested by the non-relativistic reduction of the 
one-gluon-exchange diagram~\cite{ISGU},
\begin{equation} 
H=\frac{2\alpha_s}{3m_im_j}\frac{8\pi}{3}\delta^3(\vec r)\vec S_i\cdot\vec S_j,
\label{eq9}
\end{equation} 
where $m_i$ and $\vec S_i$ are the mass and spin operator of particle $i$. As
one can see in Eqs.~\eqref{eq8} and~\eqref{eq9}, the spin-spin term is
inversely proportional to $m_im_j$ so that in the case of baryons with 
identical quarks it goes as 1/$m_i^2$. Thus, using the masses for the light
quarks $n$ ($\approx 0.3$ GeV) and for the heavy quark $b$ ($\approx 5$ GeV) 
from Refs.~\cite{BHAD,BRAC}, one gets that the spin-spin term in the case of 
the $bb$ interaction is about 250 times smaller that that of the $nn$ interaction
and therefore it is negligible. This means that the interaction between two
$b$ quarks is basically independent of the spin. 

Let us now consider the spin-spin interaction in the case of two $\Omega_{bbb}$
baryons. In a quark model description, the baryon-baryon 
interaction can be deduced from the quark-quark interaction following a 
well-known procedure~\cite{SHIM,VAGA}. In the case where the interactions are 
restricted to $S$ waves, if the quark-quark interaction is almost independent of the
spin, the baryon-baryon interaction will also be almost independent of the 
spin~\cite{SHIM}. This would imply a purely central $\Omega_{bbb}\Omega_{bbb}$ 
interaction, so that
$V_{\Omega_{bbb}\Omega_{bbb}}(^1S_0) \sim V_{\Omega_{bbb}\Omega_{bbb}}(^5S_2)$ 
and, thus, one can use the $S_i=0$ interaction of Ref.~\cite{MATH}
also for $S_i=2$.

\section{Two-channel Faddeev problem}
\label{sec4}
The explicit form of the Faddeev equations for the two-channel problem in
momentum space is also given in the Appendix. If one includes the two channels
$S_i=0$ and $S_i=2$, the spin recoupling coefficients of Eq.~\eqref{eq6} are,
\begin{eqnarray} 
\ll 0 \rp \l. 0 \rr = -\frac{1}{4} \,\,\,\,\,\, && \,\,\,\,\, 
\ll 0 \rp \l. 2 \rr = -\frac{\sqrt{5}}{4} \nonumber \\
\ll 2 \rp \l. 0 \rr = -\frac{\sqrt{5}}{4} \,\,\,\,\,\, && \,\,\,\,\, 
\ll 2 \rp \l. 2 \rr = +\frac{3}{4} \,\, .
\label{eq10}
\end{eqnarray} 
The recoupling coefficient $\ll 2 \rp \l. 2 \rr$ is positive and much larger 
than the coefficient $\ll 0 \rp \l. 0 \rr$, so that if the interaction in 
the channel $S_i=2$ is equal
to that of the channel $S_i=0$, one would expect to get a bound state.

Following the conclusion of the previous section we start by taking
$V_{\Omega_{bbb}\Omega_{bbb}}(^5S_2)=V_{\Omega_{bbb}\Omega_{bbb}}(^1S_0)$.
We find in this case that the ground state and one excited state of the
three-body $\Omega_{bbb}\Omega_{bbb}\Omega_{bbb}$ system, 
that lie at $-$393.8 MeV and $-$94.9 MeV, respectively ($-$363.9 MeV and
$-$81.0 MeV, respectively, if one includes the Coulomb interaction).

\section{Quark level $^5S_2$ $\Omega\Omega$ interaction}
\label{sec5}
\begin{figure}[t]
\includegraphics[width=.45\columnwidth]{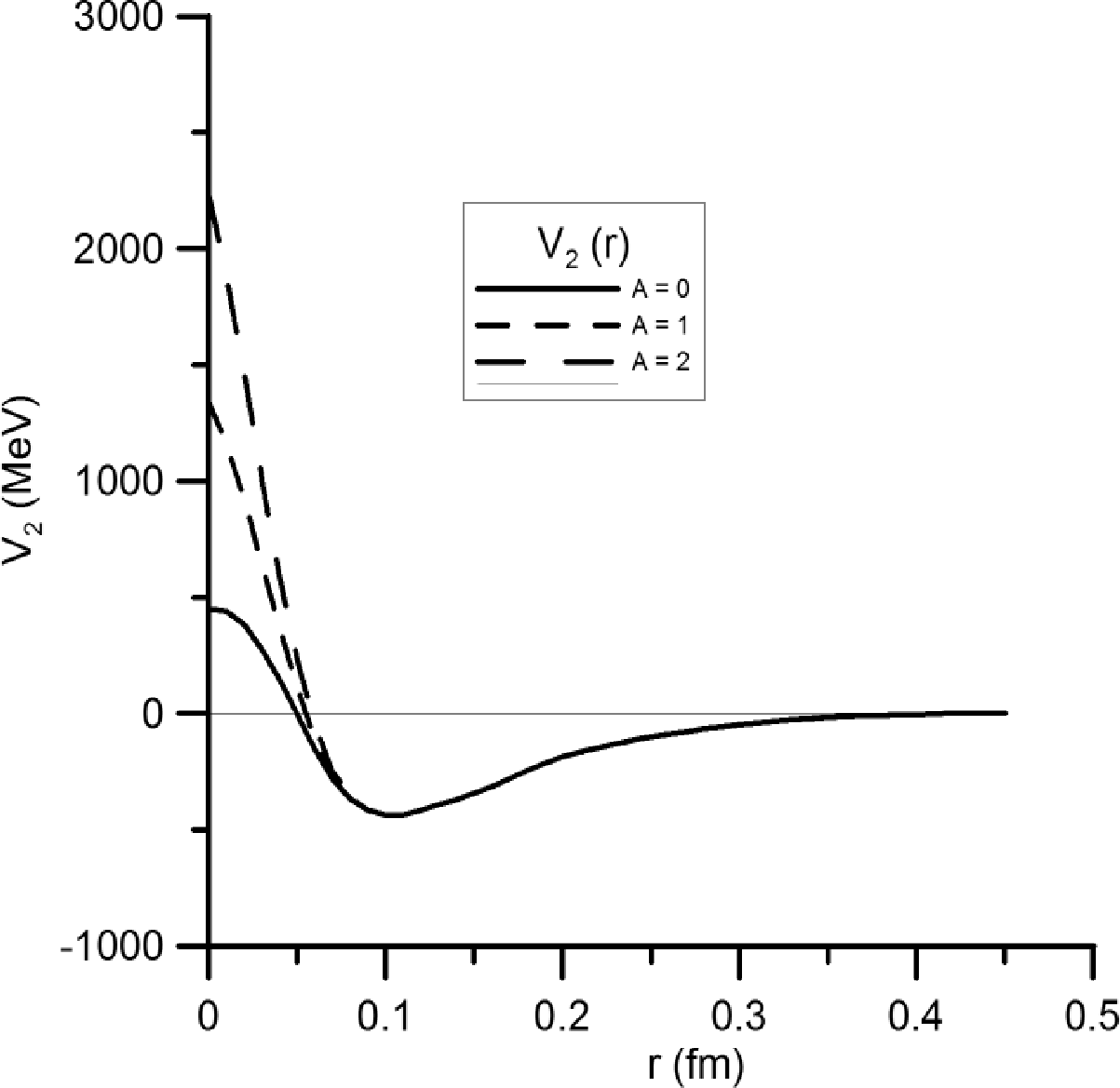}
\caption{The $\Omega_{bbb}\Omega_{bbb}$ interaction $V_2$ for $A=0,1,$ and 2.}
\label{fig1}
\end{figure}

As it has been discussed in the literature~\cite{Oka87,Oka81,Oky81,VAGA} 
there appears quark Pauli blocking for particular channels of some two-baryon
systems. This has been discussed in detail, for example, for the $\Sigma N$
system~\cite{Oka87} or for the $\Delta N$ and $\Delta\Delta$ 
systems~\cite{Oka81,Oky81,VAGA}. Pauli blocking translates into repulsive
cores of of long-range ($\sim$ 1 fm). See, for example, Fig. 3 of Ref.~\cite{Oky81}.
The $\Omega_{bbb}\Omega_{bbb}$ two-baryon belongs to the same flavor multiplet 
as the $\Delta\Delta$ system and thus it also shows Pauli blocking for the $^5S_2$ 
channel, which means a strong long-range repulsive hard core. 
In order to simulate these effects we will take the $S_i=2$ interaction as,   
\begin{equation} 
V_2(r)=\left\{ \begin{split} & V_0(r) \hspace{6cm} r > r_0 \\
& V_0(r) + A[V_0(r)-V_0(r_0)]\left(1-\frac{r} {r_0}\right)^2 \,\,\,\,\, r\le r_0
\end{split}\right. \,\,\, ,
\label{eq15}
\end{equation} 
where $A$ is a constant and $r_0=0.104$ fm is the radius where $V_0$ is
minimum (see Fig. 1). The form~\eqref{eq15} guarantees that $V_2(r)$ 
and $dV_2(r)/dr$ are continuous at $r=r_0$. This transformation has the effect
of pushing upwards the short-range part of the interaction, thereby reducing 
the  attraction. We show this behavior in Fig. 1 for the special cases
$A=1$ and $A=2$ as compared with the $A=0$ case, which corresponds to the 
model of Ref.~\cite{MATH}. Thus, our final model of the 
$\Omega_{bbb}\Omega_{bbb}$
interaction contained in Eq.~\eqref{eq15} and Fig.~\ref{fig1}, has
the $^5S_2$ interaction less attractive than the $^1S_0$ interaction which
is similar to what is obtained in  
the OBE model of Ref.~\cite{MING}.

\begin{figure}[t]
\includegraphics[width=.45\columnwidth]{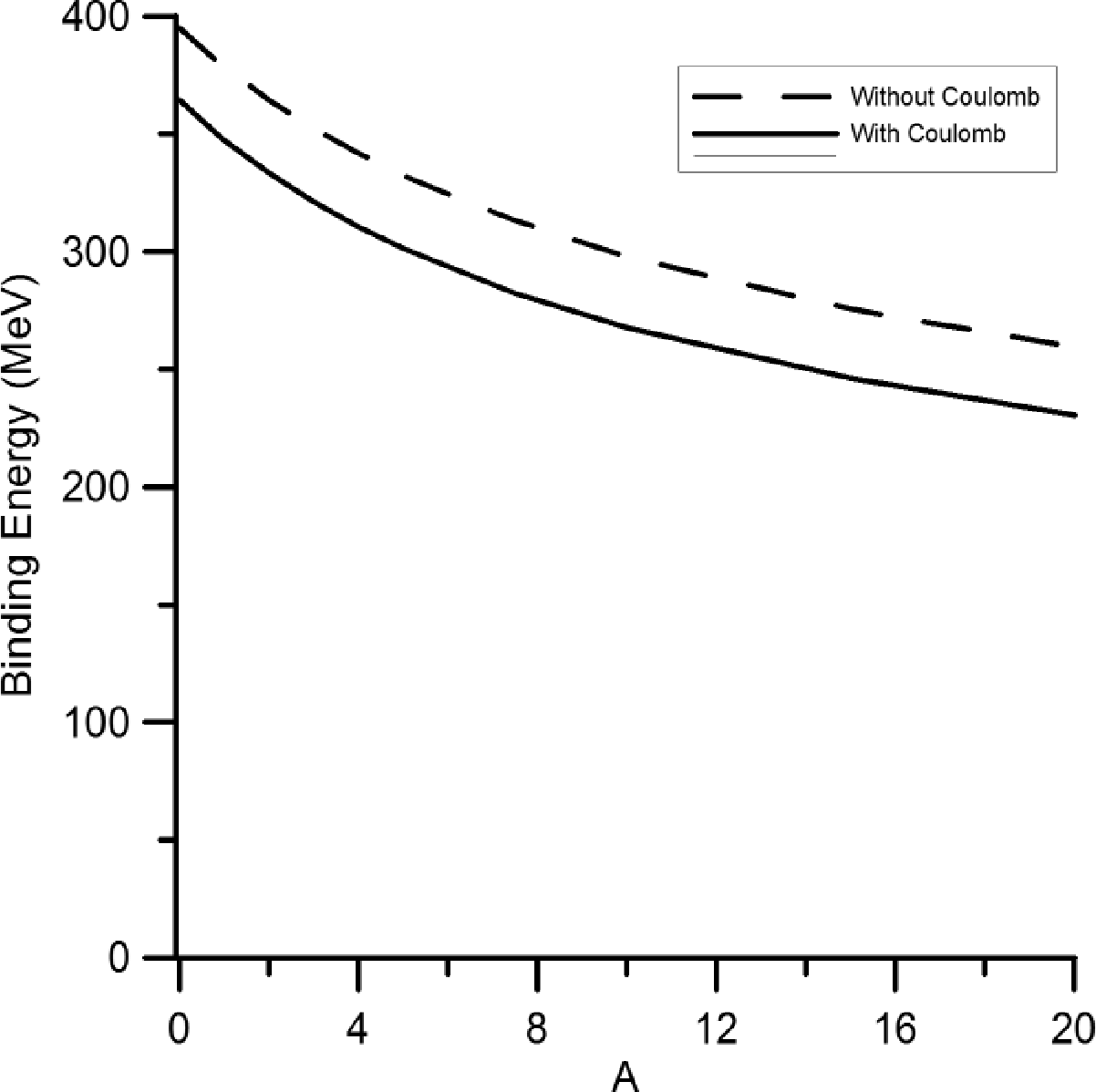}
\caption{The $\Omega_{bbb}\Omega_{bbb}\Omega_{bbb}$ binding 
energy as a function of the parameter $A$.}
\label{fig2}
\end{figure}

We show in Fig.~\ref{fig2} the evolution of the binding energy (with and without 
Coulomb) as a function of the parameter $A$. As one can see, the binding
energy changes very slowly. We consider that values of $A$ larger than 20  
are not realistic so that we expect the binding energy to lie between 
250 and 350 MeV.

\section{Acknowledgements}
This work has been partially funded by COFAA-IPN (M\'exico) and 
by the Spanish Ministerio de Ciencia e Innovaci\'on (MICINN) and the
European Regional Development Fund (ERDF) under 
contracts PID2019-105439GB-C22, PID2022-141910NB-I00,
and RED2022-134204-E.

\section{Appendix}
\label{Ape}
In order to solve the single channel three-body problem we use the method of
Ref.~\cite{GAVA}, where the two-body amplitudes are  expanded in terms of 
Legendre polynomials. Thus, the Faddeev equations for the bound state problem
take the simple form,
\begin{equation} 
T^n(q_i)=\sum_n\int_0^\infty dq_j K^{nm}(q_i,q_j;E)T^m(q_j) \,\, ,
\label{eq16}
\end{equation} 
where
\begin{equation} 
K^{nm}(q_i,q_j;E)=2\ll 0 \rp \l. 0 \rr
\sum_r\tau_i^{nr}(E-q_i^2/2\nu_i)\frac{q_i^2}{2}
\int_{-1}^1 dcos\theta\frac{P_r(x_i^\prime)P_m(x_j)}
{E-p_j^2/2\eta_j-q_j^2/2\nu_j} \,\, ,
\label{eq17}
\end{equation} 
with $\ll 0 \rp \l. 0 \rr$ the spin recoupling coefficient~\eqref{eq10}, 
\begin{equation} 
\tau_i^{nr}(e)=\frac{2n+1}{2}\frac{2r+1}{2}\int_{-1}^1dx_i\int_{-1}^1dx_i^\prime
P_n(x_i)t_i(x_i,x_i^\prime;e)P_r(x_i^\prime) \,\, ,
\label{eq18}
\end{equation} 
and
\begin{equation} 
x_i=\frac{p_i-b}{p_i+b} \,\, .
\label{eq19}
\end{equation} 
$p_j$ and $q_j$ 
are the magnitudes of the Jacobi relative momenta while $\eta_j$
and $\nu_j$ are the corresponding reduced masses.

$t_i(x_i,x_i^\prime;e)$ corresponds to the off-shell two-body t-matrix
$t_i(p_i,p_i^\prime;e)$ through the transformation~\eqref{eq19}, with $b$ a
scale parameter on which the solution does not depend. The off-shell two-body
$t$-matrices are obtained by solving the Lippmann-Schwinger equation,
\begin{equation} 
t_i(p_i,p_i^\prime;e)=V_i(p_i,p_i^\prime)+ \int_0^\infty 
{p_i^{\prime\prime}}2dp_i^{\prime\prime}
V_i(p_i,p_i^{\prime\prime})
\frac{1}{e-{p_i^{\prime\prime}}^2/2\eta_j+i\epsilon}
t_i(p_i^{\prime\prime},p_i^\prime;e) \,\, ,
\label{eq20}
\end{equation} 
with $e=E-q_i^2/2\nu_i$.
In order to solve the two-channel three-body problem, Eq.~\eqref{eq16} becomes,
\begin{equation} 
T^n_\alpha(q_i)=\sum_{m\beta}\int_0^\infty dq_j 
K^{nm}_{\alpha\beta}(q_i,q_j;E)T^m_\beta(q_j) \,\, ,
\label{eq21}
\end{equation} 
with
\begin{equation} 
K^{nm}_{\alpha\beta}(q_i,q_j;E)=2 \ll \alpha \rp \l. \beta \rr \sum_r
\tau_\alpha^{nr}(E-q_i^2/2\nu_i)\frac{q_i^2}{2}
\int_{-1}^1 dcos\theta\frac{P_r(x_i^\prime)P_m(x_j)}
{E-p_j^2/2\eta_j-q_j^2/2\nu_j} \,\, .
\label{eq22}
\end{equation}

\end{document}